\documentclass{aims} 
\usepackage{amsmath}
\usepackage{paralist}
\usepackage[misc]{ifsym}
\usepackage{epsfig} 
\usepackage{epstopdf} 
\usepackage[colorlinks=true]{hyperref}
\hypersetup{urlcolor=blue, citecolor=red}
\allowdisplaybreaks

\usepackage{bm,amssymb,amsthm}
\usepackage{subfigure}

\def\currentvolume{X}
 \def\currentissue{X}
  \def\currentyear{200X}
   \def\currentmonth{XX}
    \def\ppages{X-XX}
     \def\DOI{10.3934/xx.xxxxxxx}

\newcommand{\ini}{\mathrm{ini}}
\newcommand{\hr}{{\hat{\rho}}}
\newcommand{\hmR}{{\hat{\mathcal{R}}}}
\newcommand{\hD}{{\hat{D}}}
\newcommand{\hf}{{\hat{f}}}
\newcommand{\hg}{{\hat{g}}}
\newcommand{\hsg}{{\hat{\sigma}}}
\newcommand{\hJ}{{\hat{J}}}
\newcommand{\htt}{{\hat{t}}}
\newcommand{\hbV}{{\hat{\bm{V}}}}
\newcommand{\hV}{{\hat{V}}}
\newcommand{\bj}{{\bm{j}}}
\newcommand{\bx}{{\bm{x}}}
\newcommand{\by}{{\bm{y}}}
\newcommand{\bX}{{\bm{X}}}
\newcommand{\bxi}{{\bm{\xi}}}
\newcommand{\bz}{{\bm{\zeta}}}

\newcommand{\bv}{{\bm{v}}}
\newcommand{\id}{\mathrm{ideal}}

\newcommand{\sfg}{{\mathsf{g}}}
\newcommand{\mE}{{\mathcal{E}}}
\newcommand{\mF}{{\mathcal{F}}}
\newcommand{\mH}{{\mathcal{H}}}
\newcommand{\mR}{{\mathcal{R}}}
\newcommand{\mS}{{\mathcal{S}}}
\newcommand{\Kn}{{\mathrm{Kn}}}
\newcommand{\mHc}{{\mH^{(c)}}}
\newcommand{\mHk}{{\mH^{(k)}}}

\newcommand{\mrv}{\mathrm{v}}
\newcommand{\z}{{\zeta}}

\newtheorem{theorem}{Theorem}[section]

\theoremstyle{definition}

\newtheorem{remark}[theorem]{Remark}

\title[Numerical study of the thermal relaxation for the Enskog equation]
{Numerical study of the free energy monotonicity for the Enskog equation:
 The thermal relaxation of the dense gas between two parallel plates} 

\author[Shigeru Takata, Soma Sakata, Aoto Takahashi and Masanari Hattori]{}


\thanks{$^*$Corresponding author: Shigeru Takata}

\begin{document}
\maketitle

\centerline{\scshape
Shigeru Takata$^{{\href{mailto:takata.shigeru.4a@kyoto-u.ac.jp}{\textrm{\Letter}}}*1}$
Soma Sakata$^{{\href{mailto:sakata.soma.36k@st.kyoto-u.ac.jp}{\textrm{\Letter}}}1}$
Aoto Takahashi$^{{\href{mailto:takahashi.aoto.63c@st.kyoto-u.ac.jp}{\textrm{\Letter}}}1}$
and Masanari Hattori$^{{\href{mailto:hattori.masanari.4r@kyoto-u.ac.jp}{\textrm{\Letter}}}1}$}

\medskip

{\footnotesize
 \centerline{$^1$Department of Aeronautics and Astronautics, Kyoto University, Kyoto-daigaku-katsura, Kyoto 615-8540, Japan}
} 

%

\bigskip

 \centerline{(Communicated by Handling Editor)}


\begin{abstract}
The thermal relaxation problem between two parallel plates with the same temperature
is investigated, aiming to study the behavior of the free energy
of the dense gas described by the Enskog equation.
Two types of Enskog equation have been used:
one is the Enskog equation with the original Enskog factor,
while the other is that with a modified Enskog factor proposed recently
in Takata \& Takahashi, Phys. Rev. E \textbf{111}, 065108 (2025).
The evaluated free energy is a natural extension of the thermodynamic free energy to
the non-equilibrium state.
It is observed that this free energy monotonically decreases in time for the modified factor version, 
while it is not necessarily the case for the original version.
Differences are also observed in other quantities in the time evolution,
most typically in the density profile.
\end{abstract}



\section{Introduction\label{sec:intro}}

Recent advances in microfabrication technology have led to a growing demand for detailed analyses of gas flows in the micro-, submicro-, and even smaller systems. In such small systems, the mean free path of gas molecules is no longer sufficiently small relative to a system dimension, and the gas is not close to the local equilibrium state implicitly assumed by conventional hydrodynamic descriptions. Under these conditions, it becomes necessary to treat the gas based on microscopic considerations. Kinetic theory provides a reliable and reasonably efficient framework for such treatment, based on the Boltzmann equation and its variants \cite{C88,S07}. This paper adopts a perspective that addresses the gas behavior in such conditions in terms of the kinetic theory.

The Enskog equation \cite{E72,S16} belongs to the aforementioned framework for a dense gas. It incorporates the effects of differences in the center-of-mass positions of colliding molecules and molecular volume fraction on the intermolecular collision frequency into the Boltzmann equation,
thereby providing a kinetic description of systems even when
non-ideal gas effects manifest beyond the ideal gas regime. 
Over the past two to three decades, the Enskog equation has also been applied to the study of gas behavior in even below submicro-scale systems, 
e.g.,~\cite{F97b,F99,KKW14,WLRZ16,FGLS18,HTT22}, 
by applying the numerical method originally developed 
for the Boltzmann equation~\cite{MS96,F97a,MS97a,WZR15}. 
However, for the original Enskog equation (OEE) to which these methods have been applied, the Boltzmann H-theorem, corresponding to the second law of thermodynamics, has not been guaranteed.

The H-theorem was shown to hold not for the OEE but for the modified (or revised) Enskog equation \cite{VE73,R78}  and its adaptation to bounded domains was also discussed later in \cite{MGB18,T24}. Nevertheless, the complexity of its mathematical structure hindered the development of numerical methods for this variant. The Boltzmann--Enskog equation, which is another variant, is also known to guarantee the H-theorem. However, {this variant account for the molecular size only in the displacement of the centers of colliding molecules, not in
the change of the} collision frequency. Consequently, its use has been limited primarily to mathematical interest, { e.g., \cite{BLPT91,HN06},} not to physical applications.

In the meantime, 
we have recently shown in \cite{TT25} that the H-theorem can be guaranteed by slightly modifying a weight factor (the so-called Enskog factor) in the original Enskog equation.
This new modification is much simpler than that in \cite{VE73}.
We call this new variant the Enskog equation with a slight modification (EESM). 
The EESM can newly provide thermodynamic consistency to the numerical studies
based on the Enskog equation. 
In addition, we have also identified the contribution of the non-ideal gas effect 
that should be added to the usual Boltzmann H-function in \cite{TT25}. 
This contribution, together with the usual H-function, 
naturally extends the equilibrium entropy to non-equilibrium systems, 
and at the equilibrium state reproduces the non-ideal gas effect 
on the entropy of conventional thermodynamics.

{ Since the fluid-dynamical aspect of EESM has been investigated in \cite{TT25bull},
we will focus on the numerical demonstration of the above achievement in the present paper.
To be more specific,} 
we consider the thermal relaxation problem 
between isothermal parallel plates based on the Enskog equation, both OEE and EESM.
Since the parallel plates serve as a heat bath, the Helmholtz free energy should decrease according to the conventional thermodynamics. 
We will observe the time evolution of the free energy, which is defined based on the aforementioned extended entropy. 
For EESM, this extended free energy will be shown to decrease monotonically, both theoretically and numerically. 
However, for the OEE, actual numerical computations indicate that this is not necessarily the case.

The paper is organized as follows.
After the introduction in Sec.~\ref{sec:intro},
the problem and its formulation are presented in Sec.~\ref{sec:main};
in particular, the statement of the H-theorem for EESM \cite{TT25} is briefly summarized in Sec.~\ref{sec:H}.
Numerical method, which is a hybrid of the finite-difference method for derivatives and the fast Fourier spectral method for collision integral \cite{FMP06}, is described in Sec.~\ref{sec:Numerics}.
Numerical results are presented in Sec.~\ref{sec:results},
and the paper is concluded in Sec.~\ref{sec:Conclusion}.

\section{Problem and formulation\label{sec:main}}
\subsection{Problem\label{sec:pro}}
\begin{figure}
\centering
\includegraphics[width=0.5\textwidth]{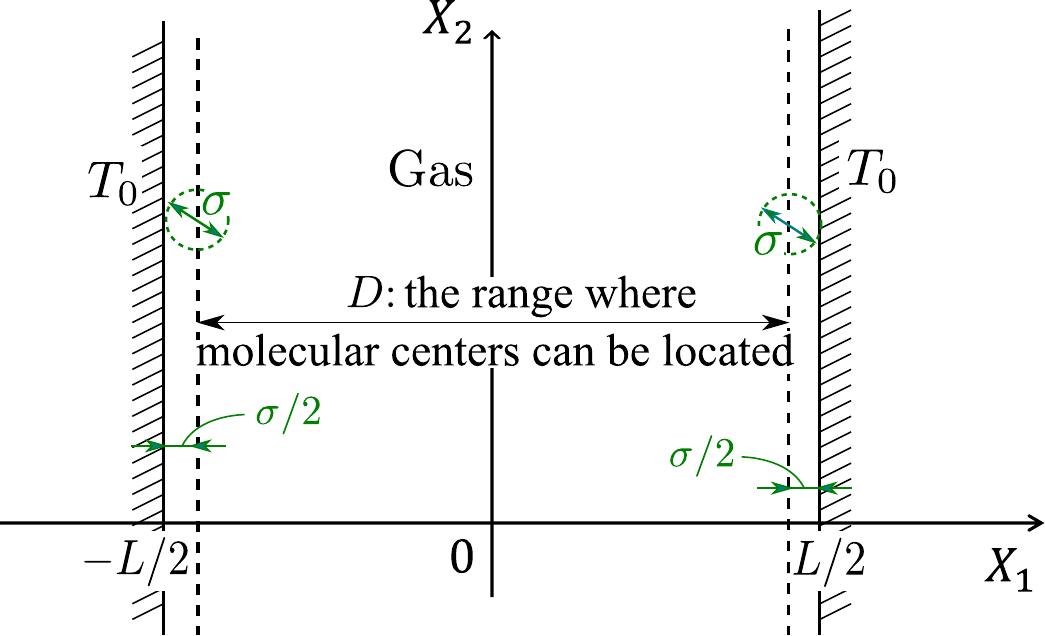}
\caption{Problem setting. Dashed circle indicates a molecule in contact with the plate surface.\label{fig:conf}}
\end{figure}

Consider a dense gas in the gap between two parallel plates at rest. 
The plates are located at $X_1=-L/2$ and $X_1=L/2$, respectively, and are maintained at a common uniform constant temperature $T_0$, see Fig.~\ref{fig:conf}. 

We will investigate the time evolution of the gas, starting from a resting Maxwellian with a uniform temperature $T_0$ and a density profile to be specified later, under the following assumptions.
\begin{enumerate}
\item The gas behavior is described by the Enskog equation. 
\item The molecules are hard-spheres with a common diameter $\sigma$ and mass $m$.
\item The gas obeys the Carnahan--Starling equation of state (Carnahan--Starling EoS) \cite{CS69}
in uniform equilibrium states.
\item The molecules are diffusely reflected \cite{S07} on the plates.
\item The gas state does not change in the directions of $X_2$ and $X_3$.
\end{enumerate}


Let $D=\{X_1| -(L-\sigma)/2 < X_1 < (L-\sigma)/2\}$, 
where the center of gas molecules can be located.
This $D$ is different from $D^\sharp=\{X_1| -L/2 < X_1 < L/2\}$,
which denotes the spatial domain between the plates.
Let $t$ and $\bm{\xi}$ be a time and a molecular velocity, respectively.
Then, denoting the one-particle velocity distribution function (VDF) of gas molecules
by $f(t,X_1,\bm{\xi}$), 
the Enskog equation is written as
\begin{subequations}\label{MEE}
\begin{align}
 & \frac{\partial f}{\partial t}+\xi_1\frac{\partial f}{\partial X_1}=J(f),
   \quad \mathrm{for\ } X_1\in D,\displaybreak[0]\label{eq:2.1}\\
 & J(f)\equiv\frac{\sigma^{2}}{m}
       \int_{\mathbb{S}^2\times\mathbb{R}^3} 
       [{g(X_1^+,X_1)f_{*}^{\prime}(X_1^+)f^{\prime}(X_1)} \notag\\
 &\qquad\qquad\quad - {g(X_1^-,X_1)f_{*}(X_1^-)f(X_1)}] \notag\\&\qquad\qquad\quad\times V_{\alpha}\theta(V_{\alpha})d\Omega(\bm{\alpha})d\bm{\xi}_{*},\label{eq:2.3}
\end{align}
\end{subequations}
\noindent
in the present problem, 
where $X_1^\pm=X_1\pm\sigma\alpha_1$,
$\bm{\alpha}=(\alpha_1,\alpha_2,\alpha_3)$ is a unit vector, 
$d\Omega(\bm{\alpha})$ is a
solid angle element in the direction of $\bm{\alpha}$,
$\theta$ is the Heaviside function
\begin{equation}
\theta(x)=\begin{cases}
1, & x\ge0\\
0, & x<0
\end{cases},
\end{equation}
\noindent
and the following notation convention has been used:
\begin{align}
 & \begin{cases}
{
f(\cdot)=f(\cdot,\bm{\xi}),\ f^{\prime}(\cdot)=f(\cdot,\bm{\xi}^{\prime})},\\
{
f_{*}(\cdot)=f(\cdot,\bm{\xi}_{*}),\ f_{*}^{\prime}(\cdot)=f(\cdot,\bm{\xi}_{*}^{\prime})},
\end{cases}\label{eq:contf}\displaybreak[0]\\
 & \begin{cases}
{
 \bm{\xi}^{\prime}=\bm{\xi}+V_{\alpha}\bm{\alpha},\quad
 \bm{\xi}_{*}^{\prime}=\bm{\xi}_{*}-V_{\alpha}\bm{\alpha},
 }\\
{ 
 V_{\alpha}=\bm{V}\cdot\bm{\alpha},\quad\bm{V}=\bm{\xi_{*}}-\bm{\xi}.}
 \end{cases}\label{eq:2.5}
\end{align}
\noindent
Here and in what follows, the argument $t$ is often suppressed, unless confusion is anticipated. 
The convention \eqref{eq:contf} will be applied only to the quantities that depend on molecular velocity.
The factor $g$ requires detailed explanation, 
which will be provided later in this subsection. 

The diffuse reflection boundary condition is written as
\begin{subequations}\label{eq:bc}%
\begin{align}
& f(\pm\frac{L-\sigma}{2})=\frac{\sigma_w^\pm}{(2\pi RT_0)^{3/2}}\exp(-\frac{\bxi^2}{2RT_0}),\quad
  \xi_1\lessgtr 0,\\
& \sigma_w^\pm=\sqrt{\frac{2\pi}{RT_0}}\int_{\xi_1\gtrless 0}|\xi_{1}|f(\pm\frac{L-\sigma}{2}) d\bxi,
\end{align}%
\end{subequations}
\noindent
while the aforementioned initial condition is
\begin{subequations}\label{eq:ic}
\begin{equation}
f=\frac{\rho_\ini(X_1)}{(2\pi RT_0)^{3/2}}\exp(-\frac{\bxi^2}{2RT_0}),
\quad t=0,
\end{equation}
\noindent
where $R$ is the specific gas constant.
In the numerical analysis, 
the following form of $\rho_\ini$ will be used mainly:
\begin{equation}
\rho_\ini(X_1)=\rho_0 (1+w\sin(\frac{2\pi}{\lambda}\frac{X_1}{L})).\label{eq:rini}
\end{equation}
\noindent
Here $w$ and $\lambda$ are positive constants, 
while $\rho_0$ is the average density defined by
\begin{equation}
\rho_0=\frac{1}{L-\sigma}\int_D \rho(X_1)dX_1,\label{eq:avrho}
\end{equation}
\noindent
with $\rho$ being the mass density:
\begin{equation}
\rho=\langle f \rangle,\label{eq:rho_def}
\end{equation}
\end{subequations}
\noindent
and $\langle\bullet\rangle\equiv\int_{\mathbb{R}^3}\bullet\ d\bxi$.
In other words, $\rho_0$ is the total mass divided by the volume of $D$ rather than $D^\sharp$.

Here some details of description are in order.
It should be noted that \eqref{MEE} makes sense only 
when the pair of positions $(X_1,X_1^\pm)$ is in $D$.
Hence, $g$ should be considered as a function
that contains the multiplications of indicator function $\chi_D$
\begin{equation}
\chi_{D}(X_1)=\begin{cases}
1, & X_1\in D\\
0, & \mbox{otherwise}
\end{cases},\label{eq:chi_def}
\end{equation}
\noindent
in a way that
\begin{equation}
g(X_1,Y_1)=\mathsf{g}(X_1,Y_1)\chi_D(X_1)\chi_D(Y_1),\label{eq:sfg}
\end{equation}%
%
\noindent
where $\mathsf{g}$ is positive and symmetric with respect to the exchange of positions: 
$\mathsf{g}(X_1,Y_1)=\mathsf{g}(Y_1,X_1)$.
We call this factor the Enskog factor in the sequel.
Although some variants of $\mathsf{g}$ have been proposed in the literature 
(e.g.,~\cite{E72,VE73,BB18,BB19}),
we focus on two variants in the present paper:
the original form by Enskog
\begin{subequations}
\begin{equation}\label{eq:gOEE}
 \mathsf{g}(X_1,Y_1)
=2\mathcal{S}(\frac{4\pi\sigma^3}{3m}\rho(\frac{X_1+Y_1}{2})),
\end{equation}
\noindent
and the form proposed in \cite{TT25}
%
\begin{equation}
 \mathsf{g}(X_1,Y_1)=\mathcal{S}(\mathcal{R}(X_1))+\mathcal{S}(\mathcal{R}(Y_1)),
\label{eq:gEESM}
\end{equation}
\noindent
with
\begin{align}
\mathcal{R}(X_1)=&\frac{2\pi}{m}\int_D \int_0^\infty\rho(Z_1)
\notag \\
&\times\theta(\sigma-\sqrt{r^2+(Z_1-X_1)^2})rdr dZ_1.
\label{eq:R_def}
\end{align}
\noindent
What we call OEE is the Enskog equation equipped with \eqref{eq:gOEE},
while what we call EESM is that equipped with \eqref{eq:gEESM}.

The form of $\mS$ is related to the equation of state of the gas
under consideration as
\begin{equation}\label{eq:EoS}
p=\rho R T (1+8\eta\mS(8\eta)),\quad \eta=\frac{\pi\sigma^3}{6m}\rho,
\end{equation}
\noindent
in the uniform equilibrium state,
where $p$ is the thermodynamic static pressure
and $T$ is the temperature defined by 
\begin{equation}
 T=\frac{1}{3R\rho}\langle (\bm{\xi}-\bm{v})^2 f \rangle, \label{eq:temp_def}
\quad \bv=\frac{1}{\rho}\langle \bxi f \rangle.
\end{equation}
\noindent
In the above, $\eta$ is the volume fraction of molecules \cite{S16}
and $\bv=(v_1,v_2,v_3)$ is the flow velocity.
Since the Carnahan--Starling EoS \cite{CS69} is written as
\begin{equation}
 \frac{p}{\rho RT}
=\frac{1+\eta+\eta^2-\eta^3}{(1-\eta)^3}
=1+\frac{2\eta(2-\eta)}{(1-\eta)^3}
,
\label{eq:C-SEoS}
\end{equation}
\noindent
the corresponding $\mS$ is given as
\begin{equation}
 \mathcal{S}(x)=\frac{16(16-x)}{(8-x)^3}. \label{eq:SCS}
\end{equation}
\end{subequations}
%

In closing this section,
it should be noted that 
in the dense gas obeying the Carnahan--Starling EoS,
the specific internal energy $e$ takes the same form
as in the case of ideal gases, namely
\begin{equation}
 e=\frac{1}{2\rho}\langle (\bm{\xi}-\bm{v})^2 f \rangle
  =\frac32RT. \label{eq:kin_inteng_def}
\end{equation}

\subsection{H-theorem\label{sec:H}}

To our best knowledge, the H-theorem has not been established for the OEE.
The EESM has been devised in \cite{TT25} in order to overcome this shortage of the OEE.
The present work primarily aims at a numerical validation of the H-theorem for EESM.

The H-theorem for EESM in \cite{TT25} is two-fold.
We restate it in a general context of a spatially one-dimensional domain $\Omega$.

(i) If $\Omega$ is isolated, most typically surrounded by a specular reflection boundary, the following function $\mH$ decreases monotonically in time:
\begin{subequations}\label{eq:mH}\begin{align}
&\mH=\mHk+\mHc, \\
&\mHk=\int_\Omega \langle f\ln (f/c_0)\rangle dX_1,\label{eq:H^k} \\
&\mHc=\int_\Omega \rho(X_1) [\int_0^{\mathcal{R}(X_1)}
\mathcal{S}(x)dx ] dX_1,\label{eq:H^c}
\end{align}\end{subequations}
\noindent
where $c_0=\rho_0(2RT_0)^{-3/2}$.\footnote{Formally, $c_0$ in \eqref{eq:H^k} can be any constant, since total mass in the physical system is conserved and its actual value does not affect the statement of the H theorem. However, for quantitative clarity, we have adopted the present definition of $c_0$ that makes the argument of the logarithmic function dimensionless.
}
Note that, in the case of the Carnahan--Starling EoS,
the integration of $\mS$ in \eqref{eq:H^c} is more specifically expressed as
\begin{equation}
 \int_0^{y} \mathcal{S}(x)dx 
=\frac{y(32-3y)}{(8-y)^2}. \label{eq:intS}
\end{equation}

(ii) If $\Omega$ is a domain contacted with an impermeable heat bath
at temperature $T_0$, the following $\mF$, instead of $\mH$, is shown to decrease monotonically in time with the aid of Darrozes--Guiraud inequality \cite{DG66}
\footnote{For this inequality to hold, the boundary conditions representing an impermeable heat bath have to satisfy a few basic physical requirements (see Appendix~A.10 of \cite{S07}). Widely used kinetic boundary conditions, such as the diffuse reflection, the Maxwell, and the Cercignani--Lampis~\cite{C88} boundary condition, satisfy these requirements.}:%
\begin{subequations}\label{eq:mF}
\begin{align}
&\mF=RT_0 \mH +\mE
    =(RT_0\mHk+\mE)+RT_0\mHc, \label{eq:mF1}\\
&\mE=\int_\Omega \rho (e+\frac12\bv^2) dX_1,\label{eq:mE}
\end{align}\end{subequations}
\noindent
where $\mE$ is the total energy.
Since the diffuse reflection boundary is a typical example of the case (ii), $\mF$, not $\mH$, ought to decrease monotonically in time in the present problem, where $\Omega$ in \eqref{eq:mH} and \eqref{eq:mF} should be read as $D=\{X_1|-(L-\sigma)/2 < X_1< (L-\sigma)/2\}$.
\noindent
%
%
\begin{remark}\label{rem1}
The $\mathcal{H}^{(c)}$ is reduced to $-\int (p-p_\id)/(RT) d\mrv$ multiplied by the total mass in the domain $\Omega$ at uniform equilibrium state at rest, where $p_\id$ is the pressure of the ideal gas and $\mrv\equiv 1/\rho$ is the specific volume. 
Since the non-ideal gas effect does not contribute to the internal energy in the Carnahan--Starling (or hard-sphere) gas, $-R\mathcal{H}^{(c)}$ is identical to the entropy attributed to the non-ideal gas effect.
As is well-known, $\mathcal{H}^{(k)}$ is identical
to the conventional H function for an ideal gas
in the Boltzmann theory, $-R\mHk$ is the ideal gas part
of the entropy in the uniform equilibrium state.
Therefore, $-R\mathcal{H}$ is a natural extension of the thermodynamic entropy $S$ 
to non-equilibrium dense gas systems.
The monotonic decrease of $\mH$ in isolated systems
is a natural consequence in view of the thermodynamic consistency.
\end{remark}
\begin{remark}\label{rem2}
Since $\mF=RT_0\mH+\mE$, it is reduced to $\mE-T_0S$
in the uniform equilibrium state (see the last three lines of Remark~\ref{rem1}). 
Therefore, $\mF$ is a natural extension of the thermodynamic Helmholtz free energy to non-equilibrium dense gas systems.
The monotonic decrease of $\mF$ is thus again a natural consequence
in view of the thermodynamic consistency.
%
Incidentally, by \eqref{eq:mF1}, $\mF-RT_0\mHc$ is none other than the ideal-gas part of the free energy.
Indeed, it is this quantity that decreases monotonically in the case of the Boltzmann equation, as shown in \cite{DG66}.
\end{remark}

\section{Numerical method\label{sec:Numerics}}

In the present paper,
we analyze the dimensionless version of the
initial- and boundary-value problem
\eqref{MEE}--\eqref{eq:ic} numerically by a finite-difference method
combined with the fast Fourier spectral method \cite{FMP06} for the collision integral.
In the dimensionless nomenclature, the problem is written as
\begin{equation}\label{eq:hMEE}
      \frac{\partial \hf}{\partial \htt}
 +\z_1\frac{\partial \hf}{\partial x_1}
 =\frac{1}{\Kn} \hJ(\hf),
\end{equation}
\noindent
with
%
\begin{align}
\hf=&\frac{2}{\pi}\big(\int_{\z_{*1}\gtrless 0} |\z_{*1}|\hf_* d\bz_*\big) \exp(-\bz^2 ), 
\notag\\
&\quad
 \mbox{for }\z_1\lessgtr 0, \quad \mbox{on }x_1=\pm \frac{1-\hsg}{2},
\label{eq:hbc}
\end{align}
\begin{equation}
\hf=[1+w\sin(\frac{2\pi}{\lambda} x_1)] \pi^{-3/2}\exp(-\bz^2), \quad \mbox{at }\htt=0,
\label{eq:hrhoin}
\end{equation}
\noindent
where $\htt$, $x_1$, $\bz=(\z_1,\z_2,\z_3)$, and $\hf$
are the dimensionless counterpart of $t$, $X_1$, $\bxi$,
and $f$, respectively; $\hat{\sigma}=\sigma/L$ and $-(1-\hsg)/2<x_1<(1-\hsg)/2$.
The details of $\hJ(\hf)$ can be found in Appendix~\ref{sec:appA}.
The $\Kn$ in \eqref{eq:hMEE} is the reference Knudsen number, defined as $\Kn=\ell_0/L$ 
with $\ell_0$ being the following reference mean free path:
\noindent
\begin{subequations}
\begin{equation}\label{eq:ell}
\ell_0=\frac{1}{\sqrt{2}\pi\sigma^2(\rho_0/m)\sfg_0},
\quad \sfg_0\equiv 2\mS(8\eta_0),\quad\eta_0=\frac{\pi\sigma^3}{6m}\rho_0.
\end{equation}
\noindent
Hence, 
$\Kn$ is expressed in terms of $\eta_0$ and $\hsg$ as
\begin{equation}\label{eq:Kn}
\Kn=\frac{\hsg}{12\sqrt{2}\eta_0\mS(8\eta_0)}.
\end{equation}
\end{subequations}
\noindent
Therefore, although the Knudsen number $\Kn$ appears as in the standard Boltzmann theory for rarefied gases, 
it can be expressed in terms of $\hsg$ and $\eta_0$, see Appendix~\ref{sec:appA} for further details.
The initial- and boundary-value problem \eqref{eq:hMEE}--\eqref{eq:hrhoin}
is thus characterized by four dimensionless parameters
$\hsg$, $\eta_0$, 
$\lambda$, and $w$.
The former two reflect the effects of the molecular size and the volume fraction\footnote{Note that, 
since $\rho_0$ is the average density based on the volume of $D$ rather than of $D^\sharp$, 
$\eta_0$ is not exactly the volume fraction at the reference state. 
The volume fraction in this state is $\eta_0(1-\sigma/L)$.},
while the latter two reflect the effects of
the wave length and amplitude of initial density variation. 

In the present paper,
the fast Fourier spectral method is adopted for the computations of collision integral. 
Accordingly, the molecular velocity space is treated 
as a periodic cubic domain with a linear length $2Z$, 
where $Z$ is a positive value such that
the original VDF is negligibly small outside of the cube $[-Z,Z]^3$.
The reader is referred to \cite{FMP06} and the references therein 
for the idea of applying the fast Fourier spectral method to the collision integral.
Its first application to the Enskog equation can be found in \cite{WZR15}.
In the present paper, we basically use the numerical code developed in \cite{HTT22} 
with proper modifications and improvements for the present problem.

\begin{table}[t]
\centering
\caption{Data of numerical discretization.\label{tab:grid}}
\begin{tabular}{lccccccccccc}
\hline\hline
      && $\Delta\htt$ & $Z$ & $N$ & $M_1$ & $M_2$ & $M_3$ && $M_\theta$ & $M_\varphi$ & $M_\hmR$\\
      \hline
G-I   && $1.0\times10^{-3}$ & $8$ & $120$ & $128$ & $8$ & $8$ && $12$ & $8$ & $16$ \\
G-II  && $2.5\times10^{-4}$ & $8$ & $480$ & $128$ & $8$ & $8$ && $12$ & $8$ & $16$ \\
G-IIIc  && $1.0\times10^{-3}$ & $8$ & $60$ & $128$ & $8$ & $8$ && $16$ & $8$ & $16$ \\
G-IIIs  && $1.0\times10^{-3}$ & $8$ & $120$ & $128$ & $8$ & $8$ && $16$ & $8$ & $16$ \\
G-IIIr  && $1.0\times10^{-3}$ & $8$ & $240$ & $128$ & $8$ & $8$ && $16$ & $8$ & $16$ \\
\hline\hline
\end{tabular}
\end{table}

Let $(\htt^{(n)},x_1^{(i)},\z_1^{(j_1)},\z_2^{(j_2)},\z_3^{(j_3)})$ denote the discretized points in time, position, and molecular velocity.
They are arranged as
\begin{align}
&
\htt^{(n)}=n\Delta\htt \quad(n=0,1,2,\dots),\\
&
x_1^{(i)}=\frac{1-\hsg}{2}\sin(\frac{\pi}{4N}i)\quad(i=-2N,\dots,2N),\\
&
\z_k^{(j_k)}=\frac{Z}{2M_k}j_k\quad(k=1,2,3; j_k=-2M_k,\dots,2M_k-1),
\end{align}
\noindent
where $\Delta\htt$ is a time step, and $N$ and $M_k$ ($k=1,2,3$) are positive integers
(see Table~\ref{tab:grid}).
We have adopted the following finite-difference approximation of \eqref{eq:hMEE}:
%
\begin{align}
&\frac{3\hf^{(n)}_{(i,\bj)}-4\hf^{(n-1)}_{(i,\bj)}+\hf^{(n-2)}_{(i,\bj)}}{2\Delta\htt}
+\z_1^{(j_1)}(\nabla_1\hf)^{(n)}_{(i,\bj)} 
\notag \\
&\qquad =\frac{1}{\Kn}
[2\hJ(\hf)^{(n-1)}_{(i,\bj)}-\hJ(\hf)^{(n-2)}_{(i,\bj)}],
\label{eq:FDS}
\end{align}
%
\noindent
where $\bj=(j_1,j_2,j_3)$, $\nabla_1\hf$ is the second-order upwind finite-difference approximation of the  $x_1$-derivative, except that it represents the first-order upwind finite-difference when $i=\pm (2N-1)$ and $j_1 \lessgtr 0$.
In the above, the following notation has also been used for conciseness:
$\Box_{(i,\bm{j})}^{(n)}=\Box(\htt^{(n)},x_1^{(i)},\z_1^{(j_1)},\z_2^{(j_2)},\z_3^{(j_3)})$
with $\Box$ being $\hf$, $\nabla_1\hf$, and $\hJ(\hf)$.

In \eqref{eq:FDS}, the second-order extrapolation of the collision term 
has been adopted on the right-hand side
to achieve the second-order accuracy in time.
When $n=1$, the discretization of the time derivative is replaced by the first-order backward approximation 
and the right-hand side is replaced by $\hJ(\hf)^{(n-1)}_{(i,\bj)}/\Kn$, forming the first-order scheme in time.
At each time step, $\hJ(\hf)^{(n)}_{(i,\bj)}$ is computed by the fast Fourier spectral method
and is stored during next two time steps.
The handling of the boundary condition also changes depending on whether
the first-order approximation or the second-order approximation is adopted.
To be specific, $\hf_*$ in \eqref{eq:hbc} is evaluated as
$2\hf_*^{(n-1)}-\hf_*^{(n-2)}$ by the Adams--Bashforth extrapolation,
similar to the right-hand side of \eqref{eq:FDS} when $n\ge2$.
For $n=1$, it is simply evaluated as $\hf_*^{(n-1)}$.
The Simpson rule is used for the integration in \eqref{eq:hbc},
while the trapezoidal rule is used for computing the moments of $\hf$. 
The Gauss--Legendre quadrature rule is used for the integration over $\bm{\alpha}$ in $\hJ$
and over $\psi$ in $\hmR$, see \eqref{eq:h2.2} and \eqref{eq:hR_def}.\footnote{The $\bm{\alpha}$ is identical to $\by/|\by|$ in the Appendix of \cite{HTT22},
where more details of computations of $\hJ$ can be found.}
The right three columns in Table~\ref{tab:grid} show the number of Legendre's zero points that are used in the latter formula.
The $M_\theta$ and $M_\varphi$ denote respectively the number of zero points 
used for the polar and the azimuth angle of $\bm{\alpha}$ 
measured from the opposite direction of $x_1$ ($x_1$ direction)
when $x_1>0$ ($x_1\le0$). 
The $M_\hmR$ is the number of zero points used for the angle $\psi$ in \eqref{eq:hR_def},
the dimensionless counterpart of \eqref{eq:R_def}.

In the computation of $\hJ$, the difference of the Enskog factor between EESM and OEE
does not cause a significant modification of the numerical code in \cite{HTT22}
and just newly requires the computation of $\hmR$, which is not computationally demanding.
The difference of EoS from \cite{HTT22}, 
which affects the function form of $\mS$, 
is also a minor issue numerically.
Therefore, the main difference from the code in \cite{HTT22} 
is the improvement to the second-order approximation in time, 
aiming at better numerical convergence in the time evolution.

Since the scheme \eqref{eq:FDS} is not of conservative structure, the total mass is not rigorously conserved. 
Thus a process of controlling total mass is preferred in the present study. 
This is realized by a simple normalization of the total mass at the end of every time step.
As the discretization is made finer, the correction made in this control process is observed to be smaller. 
It is also observed that this process does not change the obtained results essentially, 
but increases the reliability of the obtained results with the moderate size of discretization,
e.g., against the undesired influence caused by the total mass variations and 
against the influence on the judgement for time convergence to the final steady state.

After assessing the appropriate values of numerical parameters $\Delta\htt$, $Z$, $N$, and $M_1-M_3$,
the combinations in Table~\ref{tab:grid} have been adopted in the actual computations.

\section{Results\label{sec:results}}

\begin{figure}[t]
\centering
\includegraphics[width=0.72\textwidth]{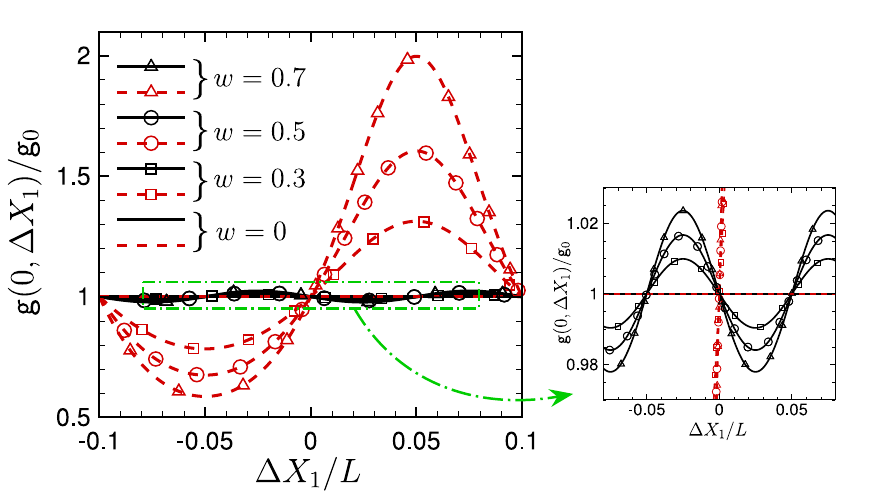}
\caption{The Enskog factor $\sfg(0,\Delta X_1)$ around the midpoint between the plates in the initial state
in the case of $\eta_0=0.25$, $\sigma/L=0.1$ ($\Kn=0.0227$), and $\lambda=0.1$ 
for $w=0$, $0.3$, $0.5$, and $0.7$. 
Solid lines: EESM. Dashed (red) lines: OEE.
The inset on the right is an enlarged view  
in the range $-0.08 \le \Delta X_1/L \le 0.08$.
\label{fig:hg}}
\end{figure}

Before carrying out the numerical computations of the problem \eqref{eq:hMEE}--\eqref{eq:hrhoin},
we have evaluated the Enskog factor $\sfg(0,\Delta X_1)$ 
for both OEE and EESM using the initial density \eqref{eq:rini}.
Here, $\Delta X_1$ is the relative position with respect to the midpoint $X_1=0$
between the plates.
If the density is uniform, the Enskog factor coincide with each other
between OEE and EESM. 
Hence, moderate variation of density does not induce a big difference.
Indeed, the results in Fig.~\ref{fig:hg} show that
the difference will be enhanced
for larger $w$.

With this preliminary observation in mind, 
we have mostly performed computations
for initial conditions with a rapid density variation, $\lambda=O(\sigma/L)$. 
Below we mainly present the results for $\eta_0=0.25$, $\sigma/L=0.1$ ($\Kn=0.0227$), and $\lambda=\sigma/L=0.1$, for which featured difference has been most clearly observed in our numerical computations.

\begin{figure}[t]
\centering
\includegraphics[width=0.5\textwidth]{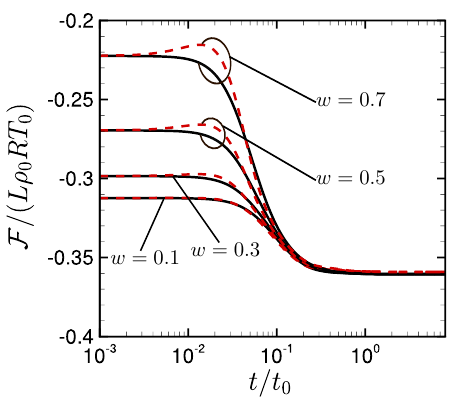}
\caption{Time evolution of $\mF$ in the case of $\eta_0=0.25$, $\sigma/L=0.1$
($\Kn=0.0227$), and $\lambda=0.1$ for $w=0.1$, $0.3$, $0.5$, and $0.7$. Solid lines: EESM. 
Dashed (red) lines: OEE. $t_0=L/\sqrt{2RT_0}$. G-I in Table~\ref{tab:grid} has been used.
\label{fig:hf1}}
\end{figure}

\subsection{Free energy\label{sec:free}}
Figure~\ref{fig:hf1} shows the time evolution of $\mF$ 
in the case of $\eta_0=0.25$, $\sigma/L=0.1$, and $\lambda=0.1$
for different values of $w$.
It is observed that $\mF$ is almost identical 
between EESM and OEE when $w=0.1$.
However, as $w$ increases, they gradually deviate from each other around between $\htt=2\times10^{-3}$ and $4\times10^{-2}$. Moreover, $\mF$ decreases monotonically in the case of EESM, but not monotonically in the case of OEE. This is a clear numerical evidence for that the H-theorem in \cite{TT25} does not hold actually in the latter case.

\begin{figure}[t]
\centering
\includegraphics[width=0.5\textwidth]{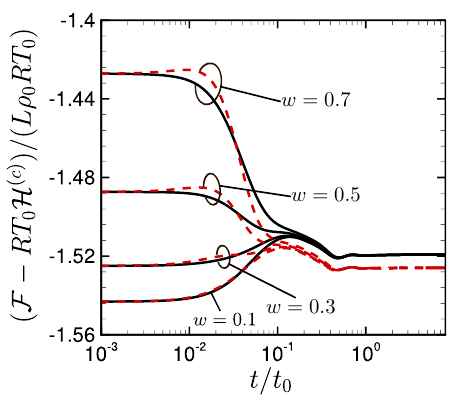}
\caption{Time evolution of the ideal gas part of free energy $\mF-RT_0\mH^{(c)}$ 
in the case of $\eta_0=0.25$, $\sigma/L=0.1$
($\Kn=0.0227$), and $\lambda=0.1$ for $w=0.1$, $0.3$, $0.5$, and $0.7$. 
$t_0=L/\sqrt{2RT_0}$. Solid lines: EESM. Dashed lines: OEE.
G-I in Table~\ref{tab:grid} has been used in computations.
\label{fig:hf2}}
\end{figure}

As a reference, the time evolution of $\mF-RT_0\mH^{(c)}$,
representing the ideal gas part of the free energy (see Remark~\ref{rem2}),
is shown in Fig.~\ref{fig:hf2}. 
As shown in this figure, this quantity does not change monotonically
either for OEE or for EESM.
This indicates that the non-kinetic contribution is crucial
for the monotonicity in the time evolution of EESM.

\begin{figure}[t]
\centering
\includegraphics[width=0.5\textwidth]{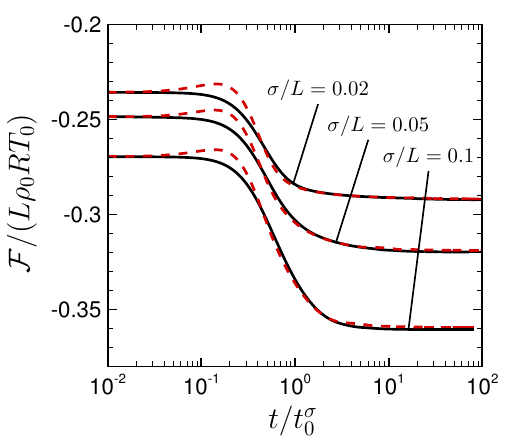}
\caption{Comparisons of the time evolution of $\mF$ for different $\sigma/L$  
in the case of $\eta_0=0.25$, $w=0.5$, and $\lambda=\sigma/L$ 
with $\sigma/L=0.02$, $0.05$, and $0.1$ (or $\Kn=0.0045$, $0.0114$, and $0.0227$). 
$t_0^\sigma=\sigma/\sqrt{2RT_0}$.
Note the time scale $t_0^\sigma$ that is different from the other figures.
Solid lines: EESM. Dashed (red) lines: OEE.
G-I (G-II) in Table~\ref{tab:grid} has been used for $\sigma/L=0.05$ and $0.1$ ($0.02$) in computations.
\label{fig:hfsig}}
\end{figure}

The influence of the molecular size relative to the gap width $L$ has also been studied. The results are shown in Fig.~\ref{fig:hfsig}. 
In this figure, the unit time scale is taken as $t_0^\sigma\equiv \sigma/\sqrt{2RT_0}$, not as $t_0=L/\sqrt{2RT_0}$,  so that the time is common for the same gas species in changing the two-plates distance $L$.
It is clearly observed that the same level of overshoot of $\mF$ for OEE occurs at the same timing.

As additional tests, the relaxation
from a uniform state, i.e., the initial data \eqref{eq:ic} with $w=0$, and
that from a concentrated state, i.e., from the initial data
\begin{equation}
\hat{f}=C[1+w_c\exp(-\lambda_c x_1^2)]\pi^{-3/2}\exp(-\bm{\zeta}^2),\quad \mbox{at }\hat{t}=0,
\label{eq:icGauss}
\end{equation}
\noindent
are studied, 
where $w_c=0.5$, $\lambda_c=1040$, and $C$ is a normalization constant
for $(1-\hat{\sigma})^{-1}\int_{\hat{D}} \hat{\rho}dx_1=1$
to hold as in the case of \eqref{eq:ic},
see \eqref{eq:hatD} in Appendix~\ref{sec:appA}.
As shown in Fig.~\ref{fig:addtest}, 
$\mathcal{F}$ monotonically decreases in both cases for EESM.
Although $\mathcal{F}$ also appears to monotonically decrease for OEE, 
a slight breakdown of monotonicity is observed numerically. 
However, these breakdowns are not definitive within the current numerical tolerance.

\begin{figure}[t]
\centering
\includegraphics[width=0.5\textwidth]{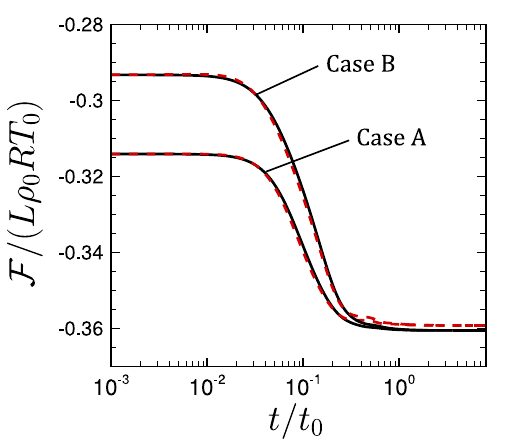}
\caption{The time evolution of $\mF$ from a uniform state,
i.e., \eqref{eq:ic} with $w=0$ (Case A), 
and that from a concentrated state,
i.e., \eqref{eq:icGauss} with $w_c=0.5$ and $\lambda_c=1040$ (Case B),
in the case of $\eta_0=0.25$ and $0.1$ (or $\Kn=0.0227$). 
$t_0=L/\sqrt{2RT_0}$.
Solid lines: EESM. Dashed (red)lines: OEE.
G-I in Table~\ref{tab:grid} has been used in computations.
\label{fig:addtest}}
\end{figure}

\begin{figure}[t]
\centering
\includegraphics[width=0.5\textwidth]{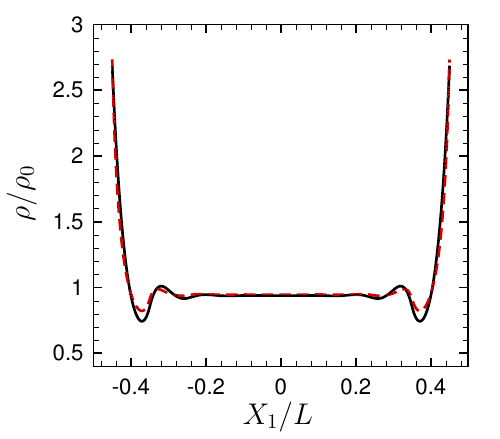}
\caption{Density profile at the final equilibrium state in the case of
$\eta_0=0.25$ and $\sigma/L=0.1$ ($\Kn=0.0227$). 
The final equilibrium state is independent of the initial condition, 
especially two parameters $\lambda$ and $w$.
Solid line: EESM. Dashed (red) line: OEE.
G-I in Table~\ref{tab:grid} has been used in computations.
\label{fig:steady}}
\end{figure}

\begin{figure}[t]
\centering
\subfigure[$t/t_0=0.01$]{\includegraphics[width=0.4\textwidth]{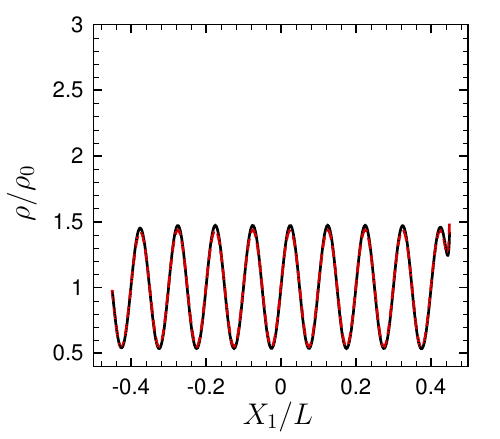}}
\subfigure[$t/t_0=0.04$]{\includegraphics[width=0.4\textwidth]{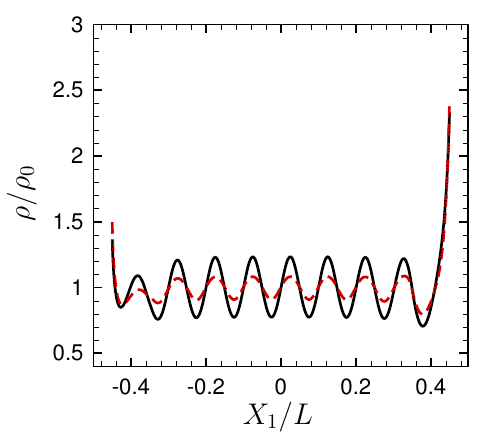}}
\subfigure[$t/t_0=0.05$]{\includegraphics[width=0.4\textwidth]{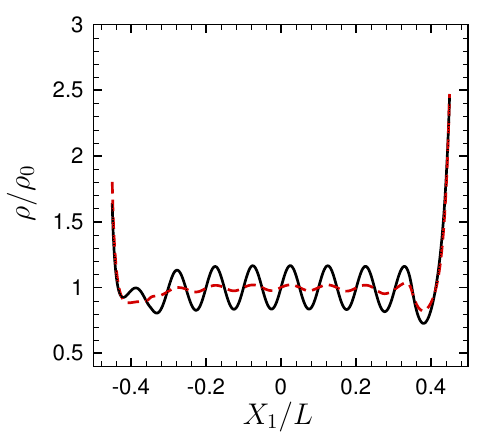}}
\subfigure[$t/t_0=0.06$]{\includegraphics[width=0.4\textwidth]{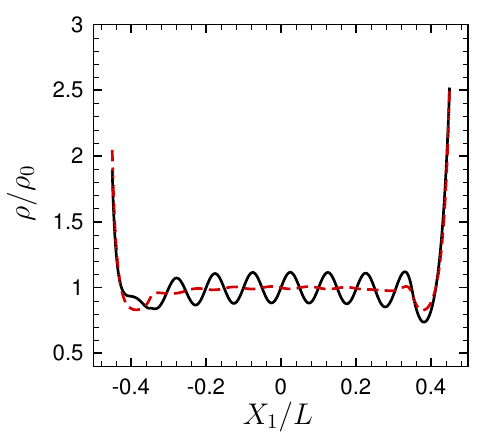}}
\subfigure[$t/t_0=0.08$]{\includegraphics[width=0.4\textwidth]{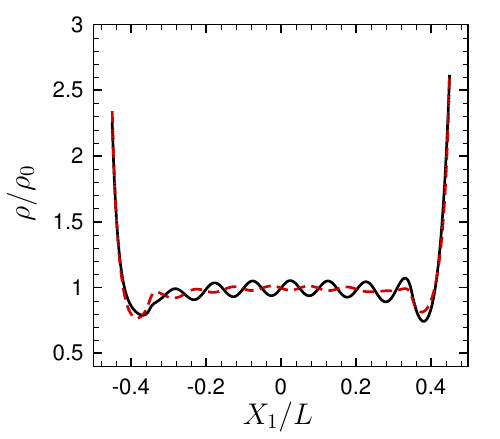}}
\subfigure[$t/t_0=0.10$]{\includegraphics[width=0.4\textwidth]{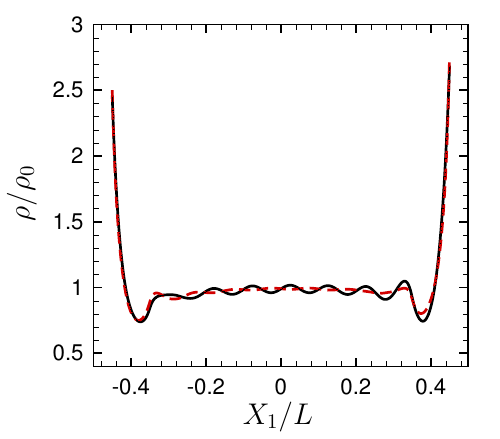}}
\caption{Time evolution of density in the case of $\eta_0=0.25$, $\sigma/L=0.1$ ($\Kn=0.0227$),
$\lambda=0.1$, and $w=0.5$. 
$t_0=L/\sqrt{2RT_0}$. Solid lines: EESM. Dashed (red) lines: OEE.
G-I in Table~\ref{tab:grid} has been used in computations.
\label{fig:dens} }
\end{figure}

Incidentally, in the actual computation, $f\ln f$ is evaluated as $|f|\ln|f|$,\footnote{In this context, $|f|\ln|f|$ should be understood as zero when $f=0$.
Similarly, $\theta(f)f\ln f$ that appears soon later should be understood as zero for $f\le0$.}
since $f$ (or $\hf$) can be negative due to the numerical error of the double precision computation.
If we use $\theta(f)f\ln f$ instead, 
the same distinct difference of time variation
between EESM and OEE are observed.
Furthermore, 
even when replacing $\mR(X_1)$ in \eqref{eq:H^c} with $[4\pi\sigma^3/(3m)]\rho(X_1)$ in accordance with \eqref{eq:gOEE},
the monotonicity has not been assured numerically for OEE.
Rather, it is sometimes observed that 
$\mF$ for OEE thus replaced may exhibit entirely different behavior from our original $\mF$ (whether OEE or EESM),
or may yield an entirely different value.  
These details are omitted here.

\begin{figure}[t]
\centering
\subfigure[$t/t_0=0.01$]{\includegraphics[width=0.4\textwidth]{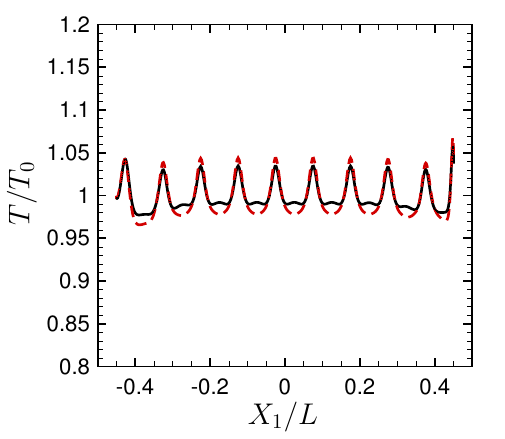}}
\subfigure[$t/t_0=0.04$]{\includegraphics[width=0.4\textwidth]{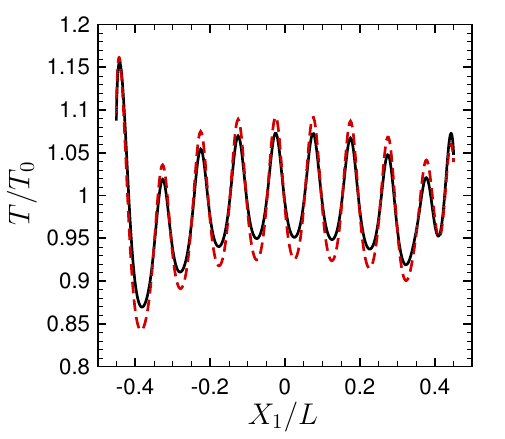}}
\subfigure[$t/t_0=0.05$]{\includegraphics[width=0.4\textwidth]{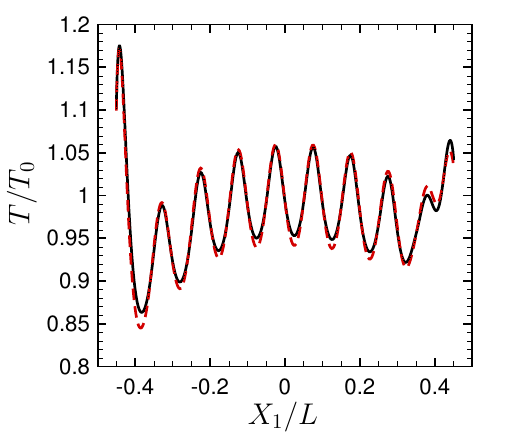}}
\subfigure[$t/t_0=0.06$]{\includegraphics[width=0.4\textwidth]{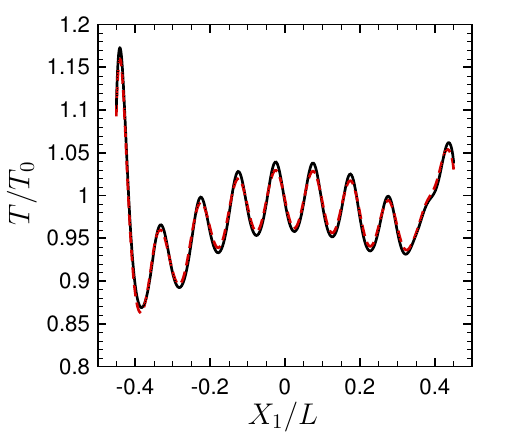}}
\subfigure[$t/t_0=0.08$]{\includegraphics[width=0.4\textwidth]{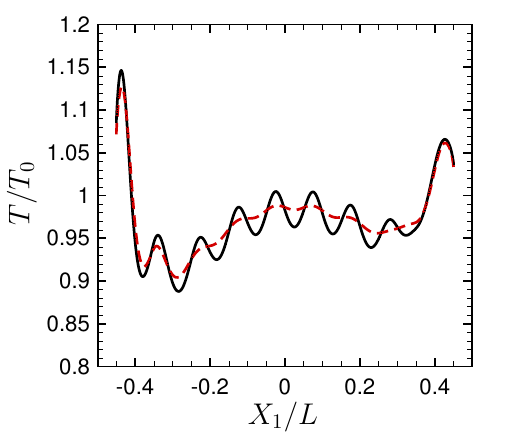}}
\subfigure[$t/t_0=0.10$]{\includegraphics[width=0.4\textwidth]{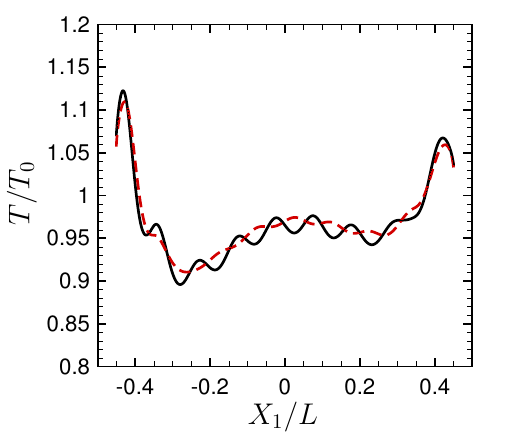}}
\caption{Time evolution of temperature in the case of $\eta_0=0.25$, $\sigma/L=0.1$ ($\Kn=0.0227$),
$\lambda=0.1$, and $w=0.5$. 
$t_0=L/\sqrt{2RT_0}$. Solid lines: EESM. Dashed (red) lines: OEE.
G-I in Table~\ref{tab:grid} has been used in computations.
\label{fig:temp}}
\end{figure}

\subsection{Density and temperature profiles\label{sec:denstemp}}

The density and temperature profiles at the final steady state
do not depend on the parameters $\lambda$ and $w$ in the initial condition. At the steady state, the temperature is uniform and is the same as the plate temperature. However, the density is not uniform. 
These features of the final equilibrium state in contact with the impermeable boundary are widely recognized in the literature, e.g., \cite{F97b,HM13,TT24}.
Figure~\ref{fig:steady} shows that
both EESM and OEE well capture the non-uniform density profile
at the final steady state. 
It is also seen that the difference between OEE and EESM is minor.

Nevertheless, as already shown in Sec.~\ref{sec:free}, 
there is a notable difference in $\mF$ between EESM and OEE. 
Hence, it is natural to observe whether 
similar differences appear in density and temperature. 
As an example, the time evolution of the density and temperature profiles in the case of $\eta_0=0.25$, $\sigma/L=0.1$, $\lambda=0.1$, and $w=0.5$ is shown in Figs.~\ref{fig:dens} and \ref{fig:temp}.
The density changes in time 
from the initial sinusoidal variation around its average value
to the profile shown in Fig.~\ref{fig:steady}.
During this transient process, the density profiles are clearly different between EESM and OEE, in particular from $\htt=0.01$ to $\htt=0.1$. 
The difference in temperature between EESM and OEE is less clear, but is still evident, particularly at $\htt=0.08$. Except for the time duration in Figs.~\ref{fig:dens} and \ref{fig:temp}, the density and temperature profiles are close to each other between EESM and OEE. The differences of density and temperature in the time duration in Figs.~\ref{fig:dens} and \ref{fig:temp} are also observed for other values of $w$. 
This holds true with a small magnitude even when the time evolution of $\mF$ is almost identical between EESM and OEE  such as the case $w=0.1$.

Because of the restriction in the equilibrium state (see, e.g., \cite{TT25,T24}), 
the temperature reaches the same uniform value as the plate temperature $T_0$ in the final steady state. 
Therefore, it is remarkable that the temperature varies both in time and space as shown in Fig.~\ref{fig:temp}, 
in spite of the fact that the temperature is initially uniform and is the same as that at the final steady state.

\begin{figure}[t]
\centering
\subfigure[density]{\includegraphics[width=0.4\textwidth]{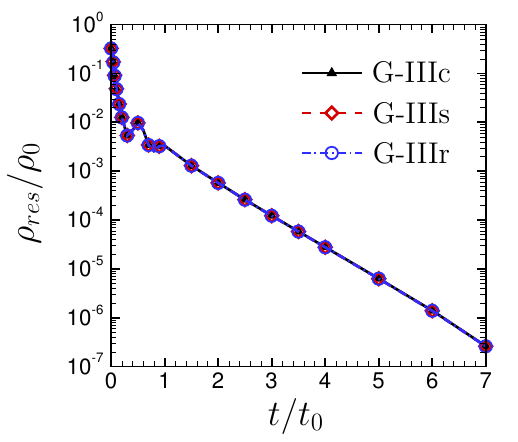}}
\subfigure[temperature]{\includegraphics[width=0.4\textwidth]{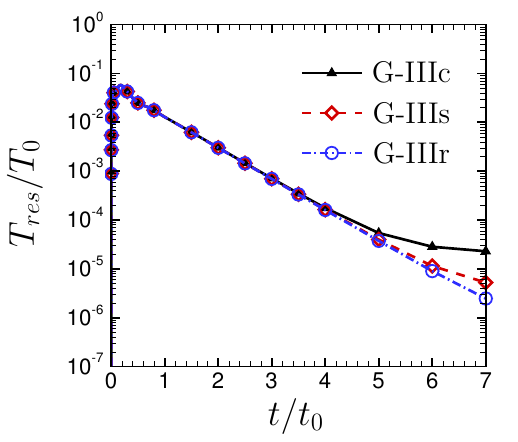}}
\caption{
The rate of approach to the final steady state 
in the case of $\eta_0=0.25$, $\sigma/L=0.1$ ($\Kn=0.0227$), $\lambda=0.1$, and $w=0.5$. 
(a) Density. (b) Temperature. 
$\rho_\mathrm{res}\equiv L^{-1}\int_D |\rho(t,X_1)-\rho_\mathrm{final}(X_1)| dX_1$,
where $\rho_\mathrm{final}$ is the density at the final steady state, while
$T_\mathrm{res}\equiv L^{-1}\int_D |T(t,X_1)-T_0|dX_1$. 
Note that the temperature at the final steady state
is uniform and equal to $T_0$. 
$t_0=L/\sqrt{2RT_0}$. Solid, dashed (red), and dash-dotted (blue) line indicate the result 
obtained by using the grids G-IIIc, G-IIIs, and G-IIIr in Table~\ref{tab:grid},
 respectively.
\label{fig:rate}}
\end{figure}

Finally, the rate of approach to the final steady state in density and temperature is shown as semi-logarithmic plots for EESM in Fig.~\ref{fig:rate}. As the grid resolution improves from G-IIIc to G-IIIs and then to G-IIIr, the saturation in the temperature plot at large time is suppressed and the linear profile widens toward the large-time regime.
Therefore, it is strongly suggested that the observed saturation in temperature is simply due to insufficient data precision and, therefore, that the approach to the steady state  is exponential rather than the algebraic both in density and temperature in the present relaxation problem.

\section{Conclusion\label{sec:Conclusion}}

In the present paper, the time relaxation toward the final equilibrium state between two parallel plates with a common temperature has been numerically studied. 
The primary aim is to observe the behavior of the free energy of the dense gas described by the Enskog equation. 
Two types of Enskog equation have been used: one is the Enskog equation with the original Enskog factor, while the other is that with a modified Enskog factor proposed recently in \cite{TT25}. 
The monitored free energy is a natural extension of the thermodynamic free energy to the non-equilibrium systems. 
It has been observed that the free energy monotonically decreases in time in the case of the modified factor and that it does not necessarily change monotonically in the case of the original factor. 
The differences of the density profiles are also observed during the time evolution, although the profiles at the final steady states are close to each other. 

The present work has thus provided the numerical evidence that supports the theoretical conclusion on the monotonic decrease of the free energy in \cite{TT25} and furthermore clear numerical counterexamples to the original Enskog equation for its H-theorem that is based on the functional $\mathcal{F}$ defined as \eqref{eq:mF}. It has further provided numerical evidence suggesting that the rate of approach to the final steady state is exponential rather than algebraic.

\section*{Acknowledgments}
The present work is supported in part by the Kyoto University Foundation, by the JSPS Grant-in-Aid (No.~24KJ1450), 
and by the HPCI system Research Project (No.~hp250004).
It is partly achieved by the use of SQUID at D3 Center, The University of Osaka.

\appendix
\section{
Dimensionless nomenclature\label{sec:appA}}

We provide a supplemental description of the dimensionless nomenclature, thereby clearly showing the independent parameters to be specified.

Let $\bx=\bX/L$, $\bz=\bm{\xi}/\sqrt{2RT_0}$,
and $\htt=t(\sqrt{2RT_0}/L)$,
which are respectively the dimensionless position vector, 
molecular velocity, and time.
Then, the Enskog equation for the dimensionless VDF $\hf(\htt,x_1,\bz)\equiv f(t,X_1,\bm{\xi})/c_0$ is obtained in the form
of \eqref{eq:hMEE} with $\hJ(\hf)$ being given by
\begin{multline}
\hJ(\hf)\equiv
\frac{1}{\sqrt{2}\pi}
\int_{\mathbb{S}^2\times\mathbb{R}^3} 
[{\hg(x_1^+,x_1)\hf_{*}^{\prime}(x_1^+)\hf^{\prime}(x_1)} \\
-{\hg(x_1^-,x_1)\hf_{*}(x_1^-)\hf(x_1)}]
  \hV_{\alpha}\theta(\hV_{\alpha})d\Omega(\bm{\alpha})d\bz_{*},
  \label{eq:h2.2}
\end{multline}
\noindent
where $x_1^\pm=x_1\pm\hsg\alpha_1$,
$\hg(x_1^\pm,x_1)=g(X_1,X_1^\pm)/\sfg_0$, and
\begin{equation} 
\begin{cases}
\bz^{\prime}=\bz+\hV_{\alpha}\bm{\alpha},\quad\bz_{*}^{\prime}=\bz_{*}-\hV_{\alpha}\bm{\alpha},\\
\hV_{\alpha}=\hbV\cdot\bm{\alpha},\quad\hbV=\bz_{*}-\bz.
\end{cases}\label{eq:h2.5}
\end{equation}

It should be noted that the specific form of $\hg$ occurring in \eqref{eq:h2.2} is different between OEE and EESM.
It is given by
\begin{subequations}
\begin{equation}
\hg(x_1,y_1)= 
\begin{cases}
2\hat{{\mS}}(\hr(\frac{x_1+y_1}{2}))\chi_\hD(x_1)\chi_\hD(y_1),\quad \mbox{(OEE),}\\
[\hat{{\mS}}(\hmR(x_1))+\hat{{\mS}}(\hmR(y_1))]\chi_\hD(x_1)\chi_\hD(y_1),\quad \mbox{(EESM),}
\end{cases}
\label{eq:hg_def}
\end{equation}
with
\noindent
\begin{equation}
\hat{D}=\{x_1| -(1-\hsg)/2<x_1<(1-\hsg)/2\},\label{eq:hatD}
\end{equation}%
\begin{equation}
 \hat{{\mS}}(\hmR(x_1))
=\frac{{\mS}({\mR}(X_1))}{\sfg_0}
=\frac12\frac{{\mS}(8\eta_0{\hmR}(x_1))}{\mS(8\eta_0)},
\label{eq:hS_def}
\end{equation}%
\begin{align}
 \hmR(x_1)
 \big(\equiv\frac{{\mR}(X_1)}{8\eta_0}\big)
=&\frac{3}{2\hsg^{3}}
 \int_{\hD} \int_0^\infty \hat{\rho}(y_1)\theta(\hsg-\sqrt{r^2+(y_1-x_1)^2}) r dr dy_1\notag\displaybreak[0]\\
=&\frac{3}{4\hsg^{3}}
 \int_{(y_1-x_1)^2\le\hsg^2} \hat{\rho}(y_1)[\hsg^2-(y_1-x_1)^2]\chi_\hD(y_1) dy_1\notag\displaybreak[0]\\
=&\frac{3}{4}
 \int_0^\pi \hat{\rho}(x_1+\hsg\cos\psi) \chi_{\hD}(x_1+\hsg\cos\psi)\sin^3\psi d\psi.
\label{eq:hR_def}
\end{align}
\end{subequations}
\noindent

The dimensionless initial condition \eqref{eq:hrhoin} and the boundary condition \eqref{eq:hbc} are obtained straightforwardly 
from the definitions of $\htt$, $\bx$, $\bz$, and $\hf$.
Then, it is clear that the details of $\hg$ do not induce parameters other than $\hsg$ and $\eta_0$. Hence, even by taking account of the two parameters $w$ and $\lambda$ in the initial condition, there are only four independent parameters in the present problem, regardless of whether it is OEE or EESM.

\medskip
Received xxxx 20xx; revised xxxx 20xx; early access xxxx 20xx.
\medskip


\begin{thebibliography}{99}
%
\bibitem{VE73}[10.1016/0031-8914(73)90372-8] 
\newblock H. van Beijeren and M. H. Ernst, 
\newblock \doititle{The modified Enskog equation}, 
\newblock \emph{Physica} \textbf{68} (1973), 437-456. 
%
\bibitem{BLPT91}
\newblock N. Bellomo, M. Lachowicz, J. Polewczak, and G. Toscani,
\newblock \emph{Mathematical Topics in Nonlinear Kinetic Theory II},
\newblock World Scientific, Singapore, 1991. 
%
\bibitem{BB18}[10.1103/PhysRevE.97.062115] 
\newblock E. S. Benilov and M. S. Benilov, 
\newblock \doititle{Energy conservation and H theorem for the Enskog--Vlasov equation}, 
\newblock \emph{Phys. Rev. E} \textbf{97} (2018), 062115. 
%
\bibitem{BB19}[10.1088/1742-5468/ab3ccf]
\newblock E. S. Benilov and M. S. Benilov,
\newblock \doititle{The Enskog--Vlasov equation: a kinetic model describing gas, liquid, and solid},
\newblock \emph{J. Stat. Mech.} \textbf{2019} (2019), 103205.
%
\bibitem{CS69}[10.1063/1.1672048]
\newblock N. F. Carnahan and K. E. Starling,
\newblock \doititle{Equation of state for non-attracting rigid spheres},
\newblock \emph{J. Chem. Phys.} \textbf{51} (1969), 635-636.
%
\bibitem{C88}[10.1007/978-1-4612-1039-9]
\newblock C. Cercignani, 
\newblock \emph{The Boltzmann Equation and Its Applications}, Springer, New York, 1988.
%
\bibitem{DG66}
\newblock J. S. Darrozes and J. P. Guiraud, 
\newblock \doititle{G\'{e}n\'{e}ralisation formelle du th\'{e}or\`{e}me H en pr\'{e}sence de parois. Applications},
\newblock \emph{C. R. Acad. Sci. Paris A} \textbf{262} (1966), 1368-1371.
%
\bibitem{E72}
\newblock D. Enskog, 
\newblock Kinetic theory of heat conduction, viscosity,
and self-diffusion in compressed gases and liquids, 
\newblock in \emph{Kinetic Theory}, Vol. 3, S. G. Brush ed., Pergamon Press, Oxford, Part 2, 1972, 226-259.
%
\bibitem{FMP06}[10.1137/050625175]
\newblock F.~Filbet, C.~Mouhot, and L.~Pareschi,
\newblock \doititle{Solving the Boltzmann equation in $N log_2N$}, 
\newblock \emph{SIAM J. Sci. Comput.} \textbf{28} (2006), 1029-1053.
%
\bibitem{F97a}[10.1063/1.869247] 
\newblock A. Frezzotti, 
\newblock \doititle{A particle scheme for the numerical solution of the Enskog equation}, 
\newblock \emph{Phys. Fluids} \textbf{9} (1997), 1329-1335. 
%
\bibitem{F97b}[10.1016/S0378-4371(97)00143-X] 
\newblock A. Frezzotti, 
\newblock \doititle{Molecular dynamics and Enskog theory calculation 
of one dimensional problems in the dynamics of dense gases}, 
\newblock \emph{Physica A} \textbf{240} (1997), 202-211.
%
\bibitem{F99}[10.1016/S0997-7546(99)80008-9]
\newblock A. Frezzotti, 
\newblock \doititle{Monte Carlo simulation of the heat flow in a dense hard sphere gas}, 
\newblock \emph{Eur. J. Mech. B/Fluids} \textbf{18} (1999), 103-119.
%
\bibitem{FGLS18}[10.1103/PhysRevFluids.3.054001]
\newblock A. Frezzotti, L. Gibelli, D. A. Lockerby, and J. E. Sprittles,
\newblock \doititle{Mean-field kinetic theory approach to evaporation of a binary liquid into vacuum},
\newblock \emph{Phys. Rev. Fluids} \textbf{3} (2018), 054001. 
%
\bibitem{HN06}[10.1088/0951-7715/19/6/001]
\newblock S.-Y. Ha and S. E. Noh,
\newblock \doititle{New \emph{a priori} estimate for the Boltzmann--Enskog equation},
\newblock \emph{Nonlinearity} \textbf{19} (2006), 1219-1232. 
%
\bibitem{HM13}[10.1016/C2010-0-66723-X]
\newblock J.-P. Hansen and I. R. McDonald,
\newblock \doititle{\emph{Theory of Simple Liquids}}, 4th ed., 
\newblock Academic Press, Oxford, 2013, Chap.~6.
%
\bibitem{HTT22}[10.1063/5.0091390]
\newblock M. Hattori, S. Tanaka, and S. Takata,
\newblock \doititle{Heat transfer in a dense gas between two parallel plates}, 
\newblock \emph{AIP Advances} \textbf{12} (2022), 055220.
%
\bibitem{KKW14}[10.1063/1.4890523]
\newblock M. Kon, K. Kobayashi, and M. Watanabe, 
\newblock \doititle{Method of determining kinetic boundary conditions in net evaporation/condensation}, 
\newblock \emph{Phys. Fluids} \textbf{26} (2014), 072003. 
%
\bibitem{MGB18}[10.1007/s10955-018-1971-7]
\newblock P. Maynar, M. I. Garcia de Soria, and J. J. Brey, 
\newblock \doititle{The Enskog equation for confined elastic hard spheres},
\newblock \emph{J. Stat. Phys.} \textbf{170} (2018), 999-1018.
%
\bibitem{MS96}[10.1103/PhysRevE.54.438]
\newblock J. M. Montanero and A. Santos,
\newblock \doititle{Monte Carlo simulation method for the Enskog equation},
\newblock \emph{Phys. Rev. E} \textbf{54} (1996), 438-444.
%
\bibitem{MS97a}[10.1063/1.869325]
\newblock J. M. Montanero and A. Santos,
\newblock \doititle{Simulation of the Enskog equation \emph{\`{a} la} Bird},
\newblock \emph{Phys. Fluids} \textbf{9} (1997), 2057-2060.
%
\bibitem{R78}[10.1007/BF01011771]
\newblock P. Resibois, 
\newblock \doititle{H-theorem for the (modified) nonlinear Enskog equation},
\newblock \emph{J. Stat. Phys.} \textbf{19} (1978), 593-609.
%
\bibitem{S07}[10.1007/978-0-8176-4573-1]
\newblock Y. Sone, 
\newblock \doititle{\emph{Molecular Gas Dynamics}}, Birkh\"{a}uer, Boston, 2007.
\newblock  Supplementary notes and errata are available from \href{http://hdl.handle.net/2433/66098}{http://hdl.handle.net/2433/66098}.
%
\bibitem{S16}[10.1093/acprof:oso/9780198716051.001.0001]
\newblock R. Soto,
\newblock \doititle{\emph{Kinetic Theory and Transport Phenomena}}, 
\newblock Oxford University Press, New York, 2016.
%
\bibitem{TT25bull}[10.21915/BIMAS.2026203]
\newblock A. Takahashi and S. Takata,
\newblock \doititle{Fluid-dynamic limit of the Enskog equation with the guaranteed H-theorem},
\newblock \emph{Bull. Inst. Math., Acad. Sin. (N.S.)} \textbf{21} (2026), 217-253.
%
\bibitem{T24}[10.3934/krm.2023025]
\newblock S. Takata, 
\newblock \doititle{On the thermal relaxation of a dense gas described by the modified Enskog equation in a closed system in contact with a heat bath},
\newblock \emph{Kinet. Relat. Mod.} \textbf{17} (2024), 331-346.
%
\bibitem{TT24}[10.3934/krm.2023040]
\newblock S. Takata and A. Takahashi,
\newblock \doititle{Note on the summational invariant and corresponding local Maxwellian for the Enskog equation},
\newblock {\emph{Kinet. Relat. Mod.}} \textbf{17} (2024), 739-754.
%
\bibitem{TT25}[10.1103/3jnf-mw6y]
\newblock S. Takata and A. Takahashi,
\newblock \doititle{Enskog and Enskog-Vlasov equations with a modified correlation factor and their H theorem},
\newblock \emph{Phys. Rev. E} \textbf{111} (2025), 065108.
%
\bibitem{WZR15}[10.1016/j.jcp.2015.09.034]
\newblock L.~Wu, Y.~Zhang, and J.~M.~Reese,
\newblock \doititle{Fast spectral solution of the generalized Enskog equation for dense gases},
\newblock \emph{J. Comput. Phys.} \textbf{303} (2015), 66-79.
%
\bibitem{WLRZ16}[10.1017/jfm.2016.173]
\newblock L. Wu, H. Liu, J. M. Reese, and Y. Zhang,
\newblock \doititle{Non-equilibrium dynamics of dense gas under tight confinement}, 
\newblock \emph{J. Fluid Mech.} \textbf{794} (2016), 252-266.
%
\end{thebibliography}
\end{document}